\documentclass[10pt,twocolumn,prX,showpacs,preprintnumbers,amsmath,amssymb,floatfix]{revtex4}
\usepackage{graphicx}
\usepackage{longtable}
\usepackage{dcolumn}
\usepackage{bm}

\begin{document}
\title{Onset of deformation at $N = 112$ in Bi nuclei}   
\author{H. Pai$^1$}
\author{G. Mukherjee$^1$}
\thanks{Corresponding author}
\email{gopal@vecc.gov.in}
\author{R. Raut$^2$}
\altaffiliation{Present address: UGC-DAE-CSR, Kolkata, India }
\author{S.K. Basu$^1$}
\author{A. Goswami$^2$}
\author{S. Chanda$^1$}
\altaffiliation{Present address: Fakir Chand College, Diamond Harbour, West Bengal, India }
\author{T. Bhattacharjee$^1$}
\author{S. Bhattacharyya$^1$}
\author{C. Bhattacharya$^1$}
\author{S. Bhattacharya$^1$}
\author{S.R. Banerjee$^1$}
\author{S. Kundu$^1$}
\author{K. Banerjee$^1$}
\author{A. Dey$^1$}
\author{T.K. Rana$^1$}
\author{J.K. Meena$^1$}
\author{D. Gupta$^1$}
\altaffiliation{Present address: Bose Institute, Salt Lake City, Kolkata 700091, India }
\author{S. Mukhopadhyay$^1$}
\author{Srijit Bhattacharya$^1$}
\altaffiliation{Present address: Barasat Govt. College, West Bengal, India }
\author{Sudeb Bhattacharya$^2$}
\author{S. Ganguly$^2$}
\altaffiliation{Present address: Chandernagore College, Hoogly, West Bengal, India }
\author{R. Kshetri$^2$}
\author{M.K. Pradhan$^2$}
\affiliation{%
$^1$Variable Energy Cyclotron Centre, 1/AF Bidhan Nagar, Kolkata 700064, INDIA\\
$^2$Saha Institute of Nuclear Physics, Kolkata 700064, INDIA}%
\author{}
\affiliation{
\\
}%
\date{\today}
\begin{abstract}

The high spin states in $^{195}$Bi has been studied by $\gamma$-ray spectroscopic method using 
the $^{181}$Ta($^{20}$Ne, 6n) fusion evaporation reaction at 130 MeV. The $\gamma\gamma$ 
coincidence data were taken using an array of 8 clover HPGe detectors. The spin and parity 
assignments of the excited states have been made from the measured directional correlation from 
oriented states (DCO) ratios and integrated polarization asymmetry (IPDCO) ratios. The results 
show, for the first time, the evidence of a rotational like band based on a 13/2$^+$ band 
head in this nucleus, indicating the onset of deformation at neutron number $N = 112$ for the 
Bismuth isotopes. The results obtained were found to be consistent with the prediction of the 
total Routhian surface calculations using Woods Saxon potential. The same calculations also 
predict a change in shape from oblate to triaxial in $^{195}$Bi at high rotational frequency.
     
\end{abstract}
       

\maketitle

\section{Introduction}
\label{sec:level01}
The nuclei near A $\sim$ 190 in the Pb region are known for rich variety of structural phenomena and interesting
shape properties. Experimental evidence of the coexistence of spherical, oblate and prolate shapes, observed in 
the lead nuclei \cite{and00,dra03} has opened up a renewed research interest in this region, both theoretically 
and experimentally. There were several spectroscopic investigations to study the shapes and single particle
level structures in the nuclei below Z = 82 shell closure \cite{be_z82}. However, there were only a few above it
\cite{pie85,lip04,nie04,mab05,har08}. 

The Nilsson diagram corresponding to this region shows that both the [505]9/2$^-$ and the [606]13/2$^+$ 
proton orbitals have strong shape driving effect towards oblate shape. The [541]1/2$^-$ and [660]1/2$^+$ 
proton orbitals, on the other hand, have strong shape driving effect towards prolate shape. The competing 
nature of different Nilsson orbitals are reflected in the calculated shapes of different nuclei in this 
region which show the occurance of various shapes including shape coexistence \cite{hey83}. Two proton 
excitations across the shell gap are generally responsible for oblate deformed structure in this region 
whereas, multiparticle excitations having four or more protons induces prolate deformations \cite{woo92}. 

In the Bismuth nuclei (Z = 83), a variety of structure is obtained from spherical \cite{lip04} to superdeformed 
shapes \cite{cla96} as one goes down in neutron number from the spherical shell closure at $N = 126$. The 
observation of rotational bands in $^{191,193}$Bi \cite{nie04} indicates a deformed shape in bismuth nuclei 
for neutron number $N =$ 109 and 110. On the other hand, the absence of any regular band like structure for 
the low lying states in the heavier odd-A Bismuth isotopes suggests spherical shapes for these nuclei at low 
excitation energy. These low lying states in $^{197-201}$Bi could be interpreted in terms of shell model and 
weak coupling of the odd proton to the neutron-hole states in the neighboring Pb core \cite{pie85,cha86}. 
Recently, a shears band has been reported at high excitation ($>$ 4 MeV) in $^{197}$Bi \cite{mab05}. The total 
Routhian surface (TRS) calculations indicate oblate deformation for this configuration. This suggests that 
deformation sets in for the multi-quasiparticle state at high excitation in $^{197}$Bi. 

In even-even Po (Z = 84) isotopes the ratio of excitation energies of 4$^+$ and 2$^+$ remains close to the 
vibrational limit ($E_{4^+}/E_{2^+} \sim 2$) until $N = 112$, below which it starts to increase towards the 
rotational limit \cite{ber95}. Similarly, the $E_{8^+}/E_{6^+}$ ratio also deviates from the limit of excitation 
energies predicted from the multi particle excitation involving $\pi h_{9/2}$ orbital ($\pi$(h$_{9/2}$)$^2$ limit) 
for the lighter isotopes with neutron number below $N = 114$. These indicate a clear evidence of structural 
change at neutron number $N \le 114$. 

The odd proton nucleus $^{195}$Bi, with neutron number $N = 112$, is an interesting transitional nucleus 
whose two immediate odd-A neighbors on either side have different shapes at low excitation energies. As 
mentioned above, the spherical shape dominates in heavier $^{197}$Bi and deformed shape dominates for 
$^{193}$Bi. The excitation energy of the 13/2$^+$ state (corresponding to $\pi i_{13/2}$ orbital), with 
respect to the 9/2$^-$ state (corresponding to $\pi h_{9/2}$ orbital), in Bismuth isotopes decreases quite 
rapidly with the decrease in neutron number. This was believed to be due to the difference between the 
interactions of $\pi i_{13/2}~-~\nu i_{13/2}$ and $\pi h_{9/2} ~-~\nu i_{13/2}$ pairs. At $N = 112$ 
the 13/2$^+$ state, originated from $\pi i_{13/2}\otimes\nu_{0^+}$ configuration, is expected to be below 
the 11/2$^-$ and 13/2$^-$ states, originated from the $\pi h_{9/2}\otimes\nu_{2^+}$ configuration. This is 
experimentally confirmed from the observation of 13/2$^+$ as the first excited state in $^{195}$Bi \cite{lon86}. 
The oblate driving nature of the proton $i_{13/2}$ orbital is expected to induce oblate deformation in $^{195}$Bi.

L\"onnroth et al. \cite{lon86} have studied the high spin states in $^{195}$Bi using $^{19}$F beam
on $^{182}$W target and $^{30}$Si beam on $^{169}$Tm target. Six $\gamma$-rays were identified in their
work. The highest state known was a 29/2$^-$, 750 ns isomer, the excitation energy of which was not 
known. Superdeformed bands have also been identified in this region in Bi isotopes in a Gammasphere 
experiment \cite{cla96} showing the rich variety of shapes one can expect in this region. However, 
no rotational like band structure has been observed at lower excitation energy in $^{195}$Bi.

\section{Experimental procedures and data analysis }
\label{sec:level02}

The $\gamma$-ray spectroscopy of $^{195}$Bi has been performed at Variable Energy Cyclotron Centre, Kolkata using 
the Indian National Gamma Array (INGA) with 8 clover HPGe detectors at the time of the experiment. The excited 
states in this nucleus were populated by fusion evaporation reaction $^{181}$Ta($^{20}$Ne,6n)$^{195}$Bi. The 
145 MeV $^{20}$Ne beam from the K130 cyclotron was degraded by about 15 MeV using a 3.6 mg/cm$^2$ Al foil 
placed 30 cm upstream to the centre of the INGA detector array. A thick (14.5 mg/cm$^2$) $^{181}$Ta target 
was used and the recoils were stopped inside the target. The clover detectors were arranged in three angles 
with respect to the beam direction. There were 2 clovers each at 40$^\circ$ and 125$^\circ$ angles while four 
clovers at 90$^\circ$. The master 
trigger was set as $\gamma - \gamma$ (and  a small run with $\gamma - \gamma - \gamma$) coincidence to collect 
list mode data. Integrated electronics modules fabricated at IUAC, New Delhi, specially for the clover detector 
pulse processing \cite{ven02} have been used in this experiment. Time and pulse height information of each 
$\gamma$-ray was stored using a CAMAC based data acquisition system. Time to Digital Converters (TDCs), used for 
individual $\gamma$-ray timing, were started by the master trigger pulse and stopped by the individual clover 
time pulse. The $\gamma - \gamma$ time difference between the $\gamma$-rays in an event ($\gamma - \gamma$ TAC) 
was obtained, in the software, by subtracting each pair of TDCs recorded in that event. A hardware Time to 
Amplitude Converter (TAC) module was also used to record the time between the master trigger and the RF signal 
of the cyclotron (RF-$\gamma$ TAC) to separately identify the ``beam-on" and ``beam-off" events. The clover 
detectors were calibrated for $\gamma$-ray energies and relative efficiencies by using $^{133}$Ba and $^{152}$Eu 
radioactive sources. 

The $\gamma - \gamma$ coincidence and intensity relations were used to build the level scheme. The data were 
sorted and analyzed using the INGASORT code \cite{bho01}. A prompt E$_\gamma$ - E$_\gamma$ matrix (matrix-1) 
was constructed gated by the prompt peak in the RF-$\gamma$ TAC and a delayed E$_\gamma$ - E$_\gamma$ matrix 
(matrix-2) was constructed gated by the delayed part (100 - 230 ns away from the prompt peak) in the same TAC. 
Both of these matrices were also gated by the prompt peak of the $\gamma - \gamma$ TAC constructed in the 
software.  

\begin{figure}[h]
\begin{center}
\includegraphics[angle = 0, scale=0.30, clip=true]{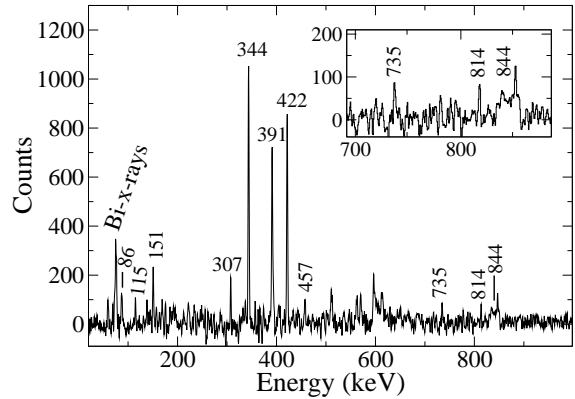}
\caption{Coincidence spectrum gated by 887-keV $\gamma$ line.}
\label{fig1}
\end{center}
\end{figure}

\begin{figure}[h]
\begin{center}
\includegraphics[angle = 0, scale=0.30, clip=true]{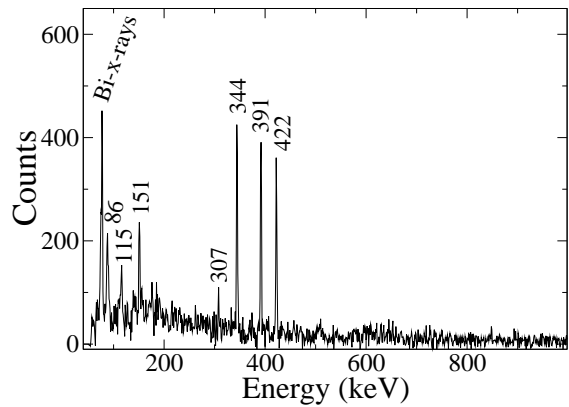}
\caption{Delayed coincidence spectrum gated by 887-keV $\gamma$ line.}
\label{fig2}
\end{center}
\end{figure}

The first excited state in $^{195}$Bi is an isomer with 32 ns halflife which decays by 887-keV $\gamma$ ray. 
There are other high spin isomers in $^{195}$Bi which were identified with halflives of 80 ns and 750 ns by 
L\"onnroth et al. \cite{lon86}. However, the time delay curves for the transitions decaying from these isomers 
were shown to contain significant amount of prompt components. The $\gamma - \gamma$ coincidence time window 
in our experiment was 400 ns wide. During the analysis, we have set a tighter coincidence gate of 100 ns, for 
constructing both the E$_{\gamma}$ - E$_{\gamma}$ matrices, which is sufficiently wide to allow coincidences 
with the 887 keV transition. Data were analyzed using both the matrices. Gates were put on the known $\gamma$-ray 
transitions in prompt matrix for obtaining the coincidence relation. 

Figure 1 shows the $\gamma-$ray spectrum gated by the known 887-keV transition in $^{195}$Bi projected from 
the matrix-1. A few new $\gamma$ lines at 86-, 307-, 457-, 735-, 814- and 844-keV have been observed in this 
spectrum. On the basis of their coincidence relations, they are placed in the level scheme. The gated spectrum 
shown in Fig. 2 is projected from the matrix-2 using the same gating transition of 887-keV. Since the 151-keV 
$\gamma$ ray is a direct decay from the 80 ns isomer, the relative intensity of this $\gamma$ line has been 
found to be higher in Fig. 2 than in Fig. 1, as expected. While the ratio of the intensities of 151- and 391-keV 
$\gamma$ rays is 0.17(4) in Fig. 1, the same ratio has been found to be 0.25(2) in Fig. 2. Interestingly, the 
ratio of intensities of the 391- and 422-keV $\gamma$ rays has also been found to be different in the two spectra. 
This has been discussed in the next section.

The multipolarities of the $\gamma$-ray transitions have been determined from the angular correlation analysis 
using the method of directional correlation from oriented states (DCO) ratios, following the prescriptions of 
Kr\"{a}mer-Flecken et al. \cite{kram89}. For the DCO ratio analysis, the coincidence events were sorted into 
an asymmetry matrix with data from 90$^\circ$ detectors ($\theta_1$) on one axis and 40$^\circ$ detectors 
($\theta_2$) on the other axis. The DCO ratios are obtained, from the $\gamma$-ray intensities (I$_\gamma$) at 
two angles $\theta_1$ and $\theta_2$, as
\begin{equation}\label{rdco}
R_{DCO} = \frac{I_{\gamma_1} ~ at~ \theta_1, ~gated ~by ~\gamma_2 ~at ~\theta_2}
               {I_{\gamma_1} ~at ~\theta_2 ~gated ~by ~\gamma_2 ~at ~\theta_1}
\end{equation}
By putting gates on the transitions with known multipolarity along the two axes of the above matrix, the
DCO ratios are obtained for each $\gamma$ ray. For stretched transitions, the value of R$_{DCO}$ would be close
to unity for the same multipolarity of $\gamma_1$ and $\gamma_2$. For different multipolarities of $\gamma_1$
and $\gamma_2$, the value of R$_{DCO}$ depends on the mixing ratio of the transitions. The value of the DCO 
ratio of a pure dipole transition, gated by a pure quadrupole transition has been calculated as 1.84 in the
above geometry. The procedure has been validated by the $\gamma$ rays of known multipolarity which were 
produced in the present experiment. R$_{DCO}$ = 1.03(7) and 0.98(4) were obtained for the known quadrupole (E2) 
transitions in $^{194}$Pb and $^{196}$Pb \cite{nds}, respectively, and R$_{DCO}$ = 1.54(23) was obtained for a 
known mixed (M1+E2) transition \cite{cha86} when gated by known quadrupole transitions. To obtain the DCO ratios 
for the $\gamma$ rays in $^{195}$Bi, the 887-keV $\gamma$ was taken as the gating transition which was known to 
be of $M2 (+E3)$ type. The measured K- conversion coefficient ($\alpha_K$) of this transition and the half-life 
of 32 ns measured for the 13/2$^+$ state \cite {lon86} indicate that the 887-keV $\gamma$ ray is, most likely, 
an M2 transition. So, the DCO values obtained for the $\gamma$ rays in $^{195}$Bi, in the present work, may be 
compared with the values calculated using a quadrupole gate. 

\begin{figure}[h]
\begin{center}
\includegraphics[angle = 0, scale=0.33, clip=true]{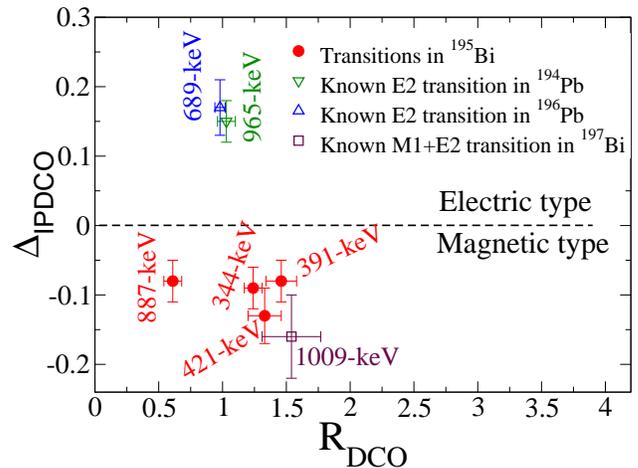}
\caption{(Color online) The DCO and the IPDCO ratios for the $\gamma$ rays in $^{195}$Bi and for a few known 
transitions in $^{194,196}$Pb and $^{197}$Bi obtained from the present work. The DCO values are obtained by 
gating on a quadrupole transition except for the 887-keV $\gamma$ ray which was gated by a dipole transition.}
\label{fig3}
\end{center}
\end{figure}

The advantage of the clover geometry of the detectors is that these detectors can be used as a $\gamma$ ray
polarimeter, using the Compton scattered events detected in the germanium crystals parallel and perpendicular 
to the directions of the reaction plane. In these measurements, one can determine the type of a $\gamma$ ray 
($E/M$) and thereby, the parity of a state can be unambiguously determined. We have measured the integrated 
polarization asymmetry (IPDCO) ratios, as described in Ref. \cite{dro96} and following the procedure
adopted in Ref. \cite{kst99}. The IPDCO asymmetry parameters have been deduced using the relation,
\begin{equation}\label{ipdco}
\Delta_{IPDCO} = \frac{a(E_\gamma) N_\perp - N_\parallel}{a(E_\gamma)N_\perp + N_\parallel},
\end{equation}
where $N_\parallel$ and $N_\perp$ are the counts for the actual Compton scattered $\gamma$ rays in the 
planes parallel and perpendicular to the reaction plane. To obtain these, two $\gamma - \gamma$ 
matrices were constructed with X-axis contained the events in which one of the $\gamma$ rays was scattered 
in the parallel (or perpendicular) direction inside the Clover detector while Y-axis contained a coincident 
$\gamma$ ray detected in any of the clover detectors. The counts were obtained from a sum gated spectrum with
gates on all low lying intense transitions. A correction factor due to the asymmetry in the array 
and response of the clover segments was incorporated which is defined by $a(E_\gamma$) = $\frac{N_\parallel} 
{N_\perp}$.  The value of this asymmetry parameter, for an ideal array, should be close to unity. In the 
present experiment this was obtained for different $\gamma$ ray energies using $^{152}$Eu source and an 
average value was found to be 1.07(5). This value compares well with the value 1.00(3) measured earlier for 
a similar array \cite{rp00}. By using the value of $a(E_\gamma$), $\Delta_{\it IPDCO}$ of the $\gamma$ rays 
in $^{195}$Bi have been determined. Positive and negative values of $\Delta_{IPDCO}$ indicate electric and 
magnetic type of transitions, respectively. The IPDCO ratios could not be obtained for the $\gamma$ rays 
having low energy (E$_\gamma \le 300$ keV) and low intensity. 

\begin{table*}
\caption{\label{tab:Table1} Energies ($E_\gamma$), intensities ($I_\gamma$), DCO ratios (R$_{DCO}$), 
IPDCO ratios ($\Delta_{IPDCO}$) and deduced multipolarities of the $\gamma$ rays in $^{195}$Bi. 
The energies of initial states ($E_i$) and spins and parities of initial ($J^\pi_i$ ) and final 
($J^\pi_f$) states are also given. }

\begin{longtable}{ccccccc}\hline

$E_{\gamma}$~~ &~~ $E_{i}$~~ &~~ $J^{\pi}_i$$\rightarrow$ $J^{\pi}_f$~~ &~~ $I_{\gamma}$
$^{\footnotemark[1]}$~~~ &~~~ $R_{DCO}$~~ &~~ $\Delta_{IPDCO}$~ &~ Deduced \\
(in keV) & (in keV) & & & (Err) & (Err) & Multipolarity \\
\hline

 86.3(2)  & 2395.8 & $29/2^{(-)} $$ \rightarrow$ $25/2^{(-)} $ & 9.3(15) & 1.04(20)$^{\footnotemark[2]}$  &    -   & E2\\

 114.9(3) & 2309.5 & $25/2^{(-)} $$ \rightarrow$ $ 23/2^{+}$ &9.7(11) & 1.41(29)$^{\footnotemark[2]}$ &   -     &E1 \\

 150.7(2) & 2194.6 & $ 23/2^{+} $$ \rightarrow$ $ 19/2^{+}$ & 10.8(9) & 1.08(20)$^{\footnotemark[2]}$ &   -    & E2  \\

 307.4(3) & 1537.8 & $ 17/2^{(+)} $$ \rightarrow$ $ 15/2^{+}$ &9.6(11)& 1.59(33)$^{\footnotemark[2]}$&   -     & (M1+E2) \\

 343.7(1) & 1230.6 & $ 15/2^{+} $$ \rightarrow$ $ 13/2^{+} $ &48.4(37)& 1.24(7)$^{\footnotemark[2]} $&  -0.09(3)   & M1+E2 \\

 391.3(2) & 1621.6 & $ 17/2^{+}  $$ \rightarrow$ $ 15/2^{+}$ &38.0(30) &1.46(12)$^{\footnotemark[2]} $& -0.08(3)    &M1+E2 \\

 421.6(1) & 2465.6 & $ (21/2^{+})  $$ \rightarrow$ $ 19/2^{+} $ &6.1(80)& 1.33(18)$^{\footnotemark[2]} $&- & (M1+E2) \\

 421.7(1) & 2043.9 & $ 19/2^{+}  $$ \rightarrow$ $ 17/2^{+} $ &37.0(80)& 1.35(19)$^{\footnotemark[4]} $&-0.13(4) & M1+E2 \\

 457.4(6) & 2923.0 & $ (23/2^{+}) $$ \rightarrow$ $ (21/2^{+})$ & 8.5(15) & 1.56(40)$^{\footnotemark[2]}$&   -     & M1+E2\\

 734.7(6) &  1621.6 & $ 17/2^{+}$$ \rightarrow$ $ 13/2^{+}$ & 6.7(13) &   -                          &  -      & (E2)  \\

 813.6(3) & 2043.9 & $19/2^{+} $$ \rightarrow$ $15/2^{+}$ & 5.7(10)   &   -                            &   -     & (E2)\\

 843.6(4) & 2465.6 & $(21/2^{+}) $$ \rightarrow$ $17/2^{+}$ & 4.4(13)   &   -                             &   -     & (E2)\\

 886.7(1) & 886.7 & $13/2^{+} $$ \rightarrow$ $ 9/2^{-}$ &100(6) & 0.61(7)$^{\footnotemark[3]}$ &  -0.08(3)     & M2 \\

\hline
\end{longtable}
\footnotetext[1]{Relative $\gamma$ ray intensities are estimated from prompt spectra and
normalized to 100 for the total intensity of 886.7 keV $\gamma$ rays.}
\footnotetext[2]{From 886.7 keV (M2) DCO gate;}
\footnotetext[3]{From 391.3 keV (M1+E2) DCO gate;}
\footnotetext[4]{From 150.7 keV (E2) DCO gate;}
\end{table*}

The spin and parity of the excited states in $^{195}$Bi has been assigned from the above two measurements. In
Fig. 3, the DCO and the IPDCO ratios are plotted for the $\gamma$ rays in $^{195}$Bi along with those for some 
of the transitions in other nuclei, produced in the same experiment and with known multipolarities. The DCO 
ratios, for all the transitions shown in Fig. 3, were obtained by gating on a known quadrupole transition 
except for the 887-keV $\gamma$ ray in $^{195}$Bi. The DCO ratio for this $\gamma$ ray was obtained from a 
dipole gate of 391-keV. As mentioned earlier, to compare our results with the known transitions, the DCO and 
IPDCO ratios obtained for the 689-, 965- and 1009-keV $\gamma$ rays, belonging to $^{196}$Pb, $^{194}$Pb and 
$^{197}$Bi, respectively, are also shown in Fig. 3 and the results are found to be consistent with the known 
type and multipolarities of the transitions \cite{nds,cha86}. 

\section{Results}
\label{subsec:level03}

The results obtained in the present work for the excited states in $^{195}$Bi are summarized in 
Table 1.  The level scheme of $^{195}$Bi, as obtained in the present work, is shown in Fig. 4. 
This level scheme includes 7 new $\gamma$ transitions over and above the ones reported earlier 
\cite{lon86}. These new lines have been placed in the level scheme and are marked by * in Fig. 4. 
The 344- and 391- keV $\gamma$-rays were assigned as dipole in character from the angular distribution 
measurements by  L\"onnroth et al. \cite{lon86}. The conversion coefficient measurements, 
performed in the same study, were not clean enough to determine their electric or magnetic 
character. In the present work, these transitions were found to be of mixed $M1 (+E2)$ in 
character from their R$_{DCO}$ and $\Delta_{IPDCO}$ values (see Fig. 3). It may be noted that the
possibility of a mixed transition was not ruled out in Ref. \cite{lon86} for the 344 keV $\gamma$ 
ray. Similarly, $E2$ multipolarity was tentatively assigned by L\"onnroth et al. for the 422 
keV transition decaying from the 2044 keV state. In the present work, however, this $\gamma$ ray 
has been found to be a $M1 (+E2)$ transition based on the DCO and the IPDCO values. Moreover, 
the weak cross-over transitions (735- and 814-keV) were also observed in the present work
and are shown in Fig. 1. 

\begin{figure}[h]
\begin{center}
\includegraphics[angle = 270, scale=0.47, angle = 0, clip=true]{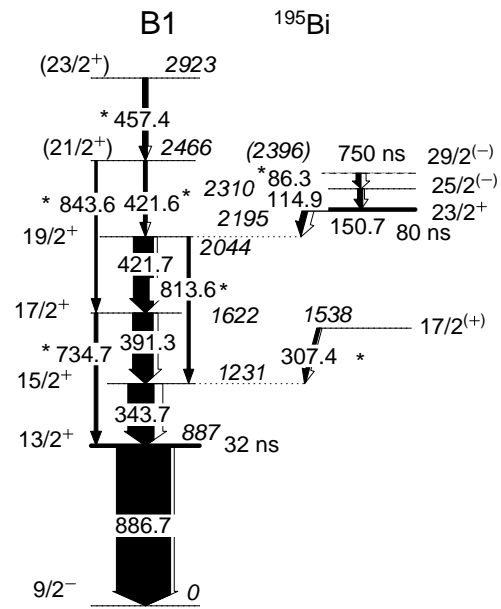}
\caption{Level scheme of $^{195}$Bi obtained from this work. The new $\gamma$ transitions are indicated by asterisks.}
\label{fig4}
\end{center}
\end{figure}

It can be seen that the intensity of 422-keV is more than the 391-keV transition in Fig. 1 whereas, it 
is less than the 391-keV transition in the delayed gate in Fig. 2. This clearly indicates the presence 
of another 422-keV prompt transition. This $\gamma$ transition has been placed above the 2044-keV state in the 
level scheme. The presence of the cross-over 844-keV line supports this placement. The intensity of the 422-keV
line has been suitably divided among the two transitions by measuring the branching ratio of the 2044-keV state
from the spectrum gated by 151-keV. A 457-keV $\gamma$ 
line is also observed in this work and was found to be in coincidence with 887-, 344-, 391-, and 422-keV
$\gamma$ rays. This $\gamma$ line was not found to be in coincidence with the 151- or 115-keV lines and
was not observed in the delayed gate. A spectrum gated by this 457-keV $\gamma$ has low statistics but
shows the indication of 844-keV line along with the other low lying transitions in the level scheme. Therefore, 
this $\gamma$ transition is placed above the 2466 keV state.  A 307-keV $\gamma$ line, observed in this work, 
has been found to be in coincidence with the 887- and 344-keV $\gamma$ rays only and has been placed accordingly.

\begin{figure}
\includegraphics[scale=0.30]{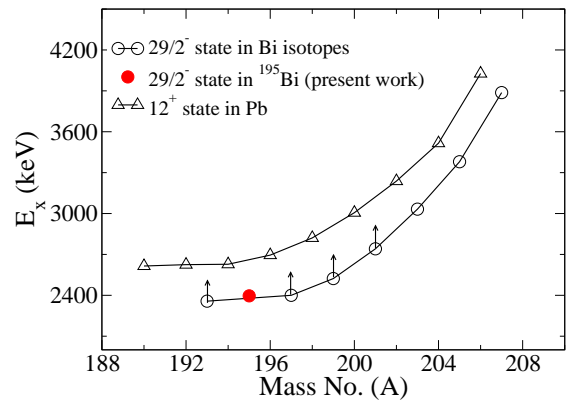}
\caption{(Color online) Systematic of the excitation energy of the 29/2$^-$ isomeric state in odd-odd Bi
nuclei (open circle) and that of the 12$^+$ state in even-even Pb nuclei (open triangle). The arrows indicate
that the values are the lower limit. The excitation energy of the 29/2$^-$ isomer in $^{195}$Bi from the present
work is shown as a solid circle.}
\label{fig5}
\end{figure}

In Ref. \cite{lon86}, no decay $\gamma$ ray was reported from the 750 ns, 29/2$^-$ isomer and hence,
the excitation energy of this isomer was uncertain. In the present work, there are indications of 
a 86-keV transition decaying from this isomer. It may be noted that this $\gamma$ line could as well 
be observed in the spectra showed in Ref. \cite{lon86} but it is difficult to assign this $\gamma$ 
ray as the K$_{\beta 1}$ X-ray of Bi has the similar energy (87.3 keV) \cite{toi}. 
From the delayed spectrum gated by 887-keV, we have obtained the ratio of intensities of the 77-keV 
($K_{\alpha1}$ of Bi) and the 86-keV as 2.66(36) whereas, the ratio of $K_{\alpha1}$ and $K_{\beta 1}$ 
X-rays (from the table in Ref. \cite{toi}) is 4.32. This suggests that there is additional 
contribution in the intensity around 86 keV. The same ratio obtained from a spectrum gated by 86-keV, 
in which the contribution of the 86-keV $\gamma$ will be absent but the contribution from the  $K_{\beta 1}$
X-ray will be present, yields a value of 4.04(80). This is in agreement with the tabulated value in Ref. 
\cite{toi}. The ratio of intensities of $K_{\alpha1}$ and $K_{\alpha2}$ was also obtained, from our data, 
as 1.77(12) which again agrees well with the value of 1.67 obtained from the table and shows that our data 
gives good estimate of the relative intensities for the pure X-rays. 

In the present work, the spin and parity of the 2310-keV state has been found to be 25/2$^-$. A 25/2$^-$ state 
has also been observed at similar excitation energy in the neighboring isotope of $^{193}$Bi \cite{nie04}. It 
is suggested that the 29/2$^-$ isomer in $^{195}$Bi decays to the 25/2$^-$ state by a 86 keV $E2$ transition. 
The measured halflife of the isomer is also consistent with the Weisskopf estimate for a 86 keV $E2$ transition. 
Therefore, the excitation energy of the 29/2$^-$ isomer is proposed to be 2396 keV. The systematic of the excitation 
energies of the 29/2$^-$ isomer in Bismuth isotopes, which arises due to the $\pi h_{9/2}\otimes \nu_{12^+}$ coupling, 
is shown in Fig. 5. The same for the 12$^+$ state in the Pb nuclei are also shown in the same Figure. It can be seen 
that the excitation energy of the 29/2$^-$ isomer in $^{195}$Bi is consistent with the systematic. 

\begin{figure}[h]
\begin{center}
\includegraphics[angle = 0, scale=0.30, angle = 0, clip=true]{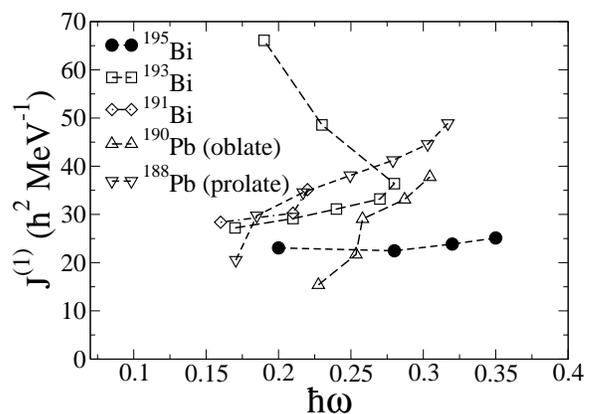}
\caption{Kinematic moments of inertai (J$^{(1)}$) as a function of rotational frequency $\hbar\omega$ for the
proposed rotational band based on 13/2$^+$ state in $^{195}$Bi along with those for neighboring odd-A bismuth
and even-even Pb isotopes.}
\label{fig6}
\end{center}
\end{figure}

\section{Discussion}
\label{sec:level03}

The band B1 in $^{195}$Bi closely resembles the rotational bands based on 13/2$^+$ band head in $^{193}$Bi and 
$^{191}$Bi \cite{nie04}. This band has the configuration of $\pi i_{13/2}$ coupled to the 2p-2h 0$^+$ intruder 
state of the Pb core with oblate deformation. Although, the $\pi i_{13/2}$ state has been observed through out the 
isotopic chain of Bi nuclei, the rotational bands based on the above configuration were observed only in the isotopes 
lighter than $A=195$, prior to the present work. Therefore, the neutron number $N = 110$ was considered to be the 
border for the observation of deformed shape in odd-A Bi isotopes. With the observation of rotational band structure 
based on the 13/2$^+$ band in $^{195}$Bi, the border has been extended to $N = 112$ in the present work. The kinematic 
moments of inertia, J$^{(1)}$, have been plotted in Fig. 6 for the above band in Bi isotopes along with those for the 
prolate and oblate structures in the Pb nuclei \cite{dra98, hee93} as in Ref. \cite{nie04} to get qualitative
information about the systematic trend of the above band structure in this region. This figure indicates that 
while the moment of inertia values in $^{191}$Bi are closer to the prolate band in $^{188}$Pb, the initial 
values in $^{195}$Bi are closer to the oblate band in $^{190}$Pb. For $^{193}$Bi, the values are in between these 
two. This indicates a gradual change in the structure of the i$_{13/2}$ band in odd-A Bi isotopes as the neutron 
number increases.

It may also be noticed in Fig. 6 that there is a band crossing in $^{193}$Bi around rotational frequency of
$\hbar\omega \sim 0.28$ MeV and there is only an indication of a band crossing in $^{191}$Bi at $\hbar\omega 
\sim 0.21$ MeV.  In the present work, however, no indication of a band crossing is observed in $^{195}$Bi
up to the highest spin observed. It indicates that there might be a delayed crossing in $^{195}$Bi which is 
reasonable from the fact that the crossing frequency increases with the neutron number as the deformation 
decreases. If one extrapolates the difference in the crossing frequencies in the two lighter isotopes, the 
crossing in $^{195}$Bi is expected at or beyond $\hbar\omega \sim 0.35$ MeV. 

The high spin single particle states in $^{193}$Bi were interpreted as due to the coupling of h$_{9/2}$ proton
with the states in the neighboring even-even $^{192}$Pb core. As mentioned earlier, the 29/2$^-$ isomer is the
result of the coupling of $\pi h_{9/2}$ with the 12$^+$ in the Pb core. The 25/2$^-$ state in $^{195}$Bi could
be interpreted as due to the coupling of the proton in the same orbital with the 8$^+$ in the Pb core in accordance
with the interpretation for the same state in $^{193}$Bi \cite{nie04}. The isomer at 2195 keV was assigned to be
a 25/2$^+$ state in $^{195}$Bi by L\"onnroth et al. \cite{lon86}. But this state was not observed in $^{193}$Bi.
The same state has been assigned to be a 23/2$^+$ state in the present work which may be interpreted as the
coupling of the $h_{9/2}$ proton with the 7$^-$ state in the Pb core. In $^{193}$Bi, the excitation energy of 
an isomer (t$_{1/2} > 10 \mu s$), observed above the 2127 keV state, suggest that this state may have the same
configuration. The 17/2$^{(+)}$ state, observed at an excitation energy of 1538 keV in this work may be interpreted
as the $\pi h_{9/2} \otimes \nu_{5^-}$ state which agrees well with the calculations shown in Ref. \cite{lon86}. 
 
\section{\bf TRS calculations}

The onset of deformation in $^{195}$Bi has been discussed in the cranking formalism. In this formalism, the total 
Routhian surface (TRS) calculations are performed using the Strutinsky shell correction method \cite{naz85,naz90}. 
Deformed Woods-Saxon potential with BCS pairing was used to calculate the single particle shell energies. The universal 
parameter set was used for the calculations. The Routhian energies were calculated in ($\beta_2, \gamma, \beta_4$) 
deformation mesh points with minimization on $\beta_4$. The procedure of such calculations have been outlined in Ref. 
\cite{muk09}. The Routhian surfaces are plotted in the conventional $\beta_2-\gamma$ plane. The TRSs are calculated for 
the different values of the rotational frequencies $\hbar\omega$ and for different configurations labelled by parity ($\pi$) 
and signature ($\alpha$) quantum numbers. At each frequency, the spin can be projected. The TRSs, calculated for the positive 
parity and $\alpha = +1/2$ configuration, are shown in Fig. 7 for the spin value of 13/2$^+$. A minimum in the TRS is obtained 
at an oblate deformation with $\beta_2 = 0.13$. This shows that the deformation driving [606]13/2$^+$ orbital, originated from 
the $\pi i_{13/2}$ level, induces oblate deformation in $^{195}$Bi. This is in qualitative agreement with the fact that the 
moment of inertia of $^{195}$Bi is closer to the oblate deformed band in $^{190}$Pb as shown in Fig. 6. Therefore, the observed 
onset of deformation in $^{195}$Bi agrees with the cranking model predictions.

\begin{figure}
\includegraphics[scale=0.30]{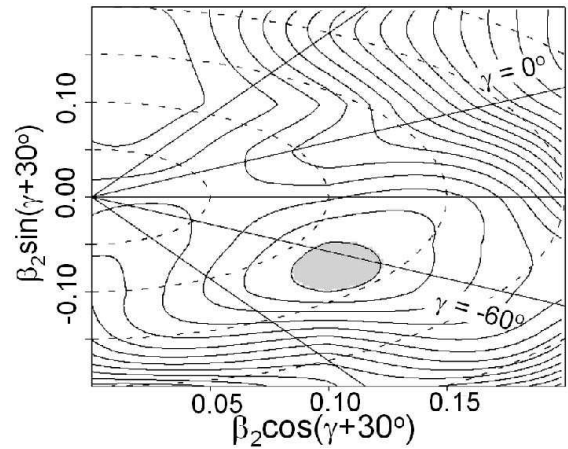}
\caption{\label{fig:fig7} Total Routhian Surfaces calculated for the 13/2$^+$ configurations in $^{195}$Bi.}
\end{figure}

The TRSs have been calculated at different rotational frequencies for the same configuration ($\pi i_{13/2}$)
in $^{195}$Bi and the values of the deformation parameters ($\beta_2$ and $\gamma$) corresponding to the minimum 
at each frequency is plotted in Fig. 8. In this plot, the $\gamma = 60^\circ$ and $\gamma = 0^\circ$ lines correspond 
to oblate and to prolate deformations, respectively. It can be seen that the calculated shape remains oblate below 
$\hbar\omega = 400$ keV after which it starts to deviate from oblate towards triaxial deformation.

\begin{figure}
\vskip 0.5 cm
\includegraphics[scale=0.28]{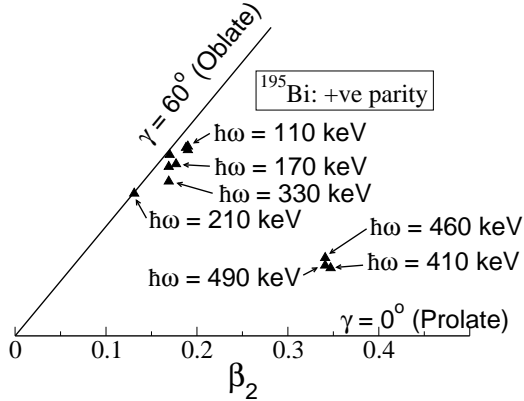}
\caption{\label{fig:fig8} Minima in the TRSs calculated for the $\pi i_{13/2}$ configuration in $^{195}$Bi
for different values of rotational frequencies $\hbar\omega$.}
\end{figure}

\section{\bf Conclusion}

The $\gamma$ ray spectroscopy of the high spin states in $^{195}$Bi has been studied using the fusion evaporation reaction
with $^{20}$Ne beam on a $^{181}$Ta target and using INGA array with 8 clover HPGe detectors.  A new level scheme with 7 
new $gamma$ transitions has been proposed for $^{195}$Bi. A rotational band based on a 13/2$^+$ band head has been proposed in 
$^{195}$Bi, similar to those observed in the lighter isotopes $^{191,193}$Bi. This indicates that onset of deformation 
takes place in the isotopic chain of Bi nuclei at $N = 112$. The excitation energy of 2396 keV proposed for the 29/2$^-$ 
isomer has been found to be consistent with the systematic of the energy of this isomer in the neighboring odd-A Bi 
isotopes. The TRS calculations using Woods-Saxon potential show oblate deformation for the 13/2$^+$ configurations 
in $^{195}$Bi. The same calculations at higher angular frequencies predict a change in shape from oblate to triaxial 
deformation around the rotational frequency of $\hbar\omega = 0.4$ MeV. More experimental work are needed to test this 
prediction.

\noindent
\section{\bf Acknowledgement}

The untiring effort of cyclotron operators at VECC are acknowledged for providing a very good beam of $^{20}$Ne. 
Acknowledgement is due to Mr. P. Mukherjee, S. Mukherjee, A. Chowdhury and late R.S. Hela for their help during 
the experiment. The help of Mr. P.K. Das is acknowledged for target and degrader foil preparation. Acknowledgements 
are also due to the electronics group of IUAC, New Delhi and Dr. A. Chatterjee and his colleagues of BARC, Mumbai 
for INGA electronics modules and multicrate data acquisition operation respectively. Fruitful discussion with Prof. 
P.M. Walker is greatefully acknowledged.\\

\end{document}